\documentclass[journal=apchd5,manuscript=article]{achemso}
\usepackage[utf8]{inputenc}
\usepackage[T1]{fontenc}

\usepackage{graphicx}
\usepackage{amsmath}
\usepackage[dvipsnames]{xcolor}
\usepackage[numbers,sort&compress,super]{natbib}
\usepackage{caption}
\usepackage[colorlinks,allcolors=blue]{hyperref}
\usepackage{bm}

\usepackage{mathtools}
\setkeys{acs}{articletitle = true}

\newcommand{\Veff}{V_{\mathrm{eff}}}
\newcommand{\Hch}{H_{\mathrm{ch}}}
\newcommand{\HEM}{H_{\mathrm{EM}}}
\newcommand{\HcE}{H_{\mathrm{ch-EM}}}
\newcommand{\Hmol}{H_{\mathrm{mol}}}
\newcommand{\Hcav}{H_{\mathrm{cav}}}
\newcommand{\Hmc}{H_{\mathrm{mol-cav}}}

\newcommand{\mb}[1]{\ensuremath{\mathbf{{#1}}}}

\title{Theoretical Challenges in Polaritonic Chemistry}
\author{J. Fregoni}
\affiliation{Departamento de Física Teórica de la Materia Condensada and Condensed Matter Physics Center (IFIMAC), Universidad Autónoma de Madrid, 28049 Madrid, Spain.}
\author{F. J. Garcia-Vidal}
\affiliation{Departamento de Física Teórica de la Materia Condensada and Condensed Matter Physics Center (IFIMAC), Universidad Autónoma de Madrid, 28049 Madrid, Spain.}
\email{fj.garcia@uam.es}
\author{J. Feist}
\affiliation{Departamento de Física Teórica de la Materia Condensada and Condensed Matter Physics Center (IFIMAC), Universidad Autónoma de Madrid, 28049 Madrid, Spain.}
\email{johannes.feist@uam.es}

\begin{document}

\begin{abstract}
Polaritonic chemistry exploits strong light-matter coupling between molecules
and confined electromagnetic field modes to enable new chemical reactivities. In
systems displaying this functionality, the choice of the cavity determines both
the confinement of the electromagnetic field and the number of molecules that
are involved in the process. Whereas in wavelength-scale optical cavities
light-matter interaction is ruled by collective effects, plasmonic subwavelength
nanocavities allow even single molecules to reach strong coupling. Due to these
very distinct situations, a multiscale theoretical toolbox is then required to
explore the rich phenomenology of polaritonic chemistry. Within this framework,
each component of the system (molecules and electromagnetic modes) needs to be
treated in sufficient detail to obtain reliable results. Starting from the very
general aspects of light-molecule interactions in typical experimental setups,
we underline the basic concepts that should be taken into account when operating
in this new area of research. Building on these considerations, we then provide
a map of the theoretical tools already available to tackle chemical applications
of molecular polaritons at different scales. Throughout the discussion, we draw
attention to both the successes and the challenges still ahead in the
theoretical description of polaritonic chemistry.
\end{abstract}

\maketitle

\begin{spacing}{1}

\section{Introduction}
Ever since the invention of the first lasers\cite{Hecht2010}, the role of light
in modern chemistry has been to act either as a probe or as a trigger to
respectively explore and induce photophysical and photochemical events. Over the
last years, a complementary paradigm based on the use of confined light modes in
micro- and nanocavities has been developed. Here, the confinement enhances the
interaction between the quantum states of light and the molecular transitions to
such an extent that the so-called strong-coupling regime is entered and the
excited states of the system become hybrids between light and matter, known as
polaritons. Polaritons inherit properties from both their constituents and also
possess new properties due to their hybrid nature, leading to significant
changes in the photophysics and photochemistry of the coupled systems. The
interest in strong coupling for modifying chemistry arose almost a decade ago
after a seminal experiment showed that photochemical reaction rates can be
modified in cavities\cite{Schwartz2011,Hutchison2012}. 

This new direction to modify and control the properties of molecular systems is
nowadays known as polaritonic chemistry\cite{Feist2018, Hertzog2019,
Garcia-Vidal2021}. It has been shown to affect a wide range of processes, such
as photochemical reactions both in single-molecule\cite{Kowalewski2016Cavity,
Galego2015, Fregoni2018, Fregoni2020Strong, Felicetti2020, Antoniou2020,
Davidsson2020Simulating, Torres-Sanchez2021} and collective\cite{Schwartz2011,
Hutchison2012, Herrera2016, Galego2016, Galego2017, Munkhbat2018, Peters2019,
Mauro2021} strong-coupling setups, as well as (possibly long-range) energy
transfer\cite{Coles2014, Georgiou2018Control, Zhong2017, Garcia-Vidal2017,
Du2018Theory, Saez-Blazquez2018Organic, Groenhof2019, Saez-Blazquez2019,
Tichauer2021, Rozenman2018, Satapathy2021}, and transitions between different
spin multiplets\cite{Stranius2018, Eizner2019, Berghuis2019, Polak2020,
Martinez-Martinez2018Polariton, Yu2021Barrier-free, Ye2021, Climent2021Notdark},
among others. We emphasize that polaritonic chemistry is not a mere substitute
of traditional chemistry techniques, as it can enable processes that are not
possible in bare materials due to the long-range and collective nature of the
polaritons.

Despite the attractiveness of these applications and the large range of existing
works, there are many open questions and fundamental problems that remain to be
addressed. The goal of this perspective is to provide an overview of and guide
through the challenges facing theoretical treatments of polaritonic chemistry,
which we hope will be useful as a guide both for scientists active in the field
and those entering it. Fundamentally, these challenges are due to the large
complexity of the studied systems, which manifests on multiple scales: the
building blocks are (often organic) molecules, which locally interact with their
environment and each other, as well as electromagnetic (EM) field modes that are
usually highly lossy and possess complex mode structures. Both of these building
blocks can be treated in arbitrary detail and possess a rich
phenomenology. Consequently, the study of each such type of subsystem in
isolation is the topic of a large field of science (respectively, chemistry and
(nano)photonics). Within polaritonic chemistry, these building blocks are made
to interact strongly, and the resulting hybrid states, the polaritons, possess
properties that are not found in either subsystem in isolation. Furthermore, in
most experimentally relevant setups, there are important collective effects,
with macroscopic numbers of molecules coupling to every single EM mode, and at
the same time, many EM modes being involved. Finally, the quantized nature of
the EM fields often plays a major role, requiring the use of techniques from
(cavity) quantum electrodynamics and quantum optics to achieve a faithful
description of the systems. Due to the often highly lossy nature of the EM
modes, these techniques often have to be combined with those of open quantum
systems.

The very general considerations above already imply that a full theoretical ab
initio modelling of such systems is effectively impossible without significant
approximations. The challenge thus lies in choosing the appropriate
simplifications and approximations in each specific situation. At the same time,
the huge available design space implies that the existing work up to now has
only scratched the surface of what is possible, and there is considerable
potential for future advancements. In order to maintain a manageable scope, in
the current perspective we focus on ``chemical'' applications, i.e., the
treatment of (collections) of molecules in the presence of quantized EM modes,
without discussing in detail how to obtain or design such modes, or uses of the
coupled systems for photonics applications. Furthermore, we restrict ourselves
for the most part to the situation where electronic transitions in the molecules
are coupled to light modes. Recent years have also seen an explosion of activity
in vibrational strong coupling, where (IR-active) transitions between
vibrational states in the molecules are coupled to cavity modes. Several recent
perspectives and reviews have treated such setups, and we encourage the
interested reader to consult those\cite{Yuen-Zhou2021, Herrera2020,
Climent2021Cavity, Garcia-Vidal2021, Simpkins2021, Wang2021Roadmap,
Nagarajan2021}.

\section{Overview of experimental setups}

\begin{figure}[tbh]
	\includegraphics[width=\linewidth]{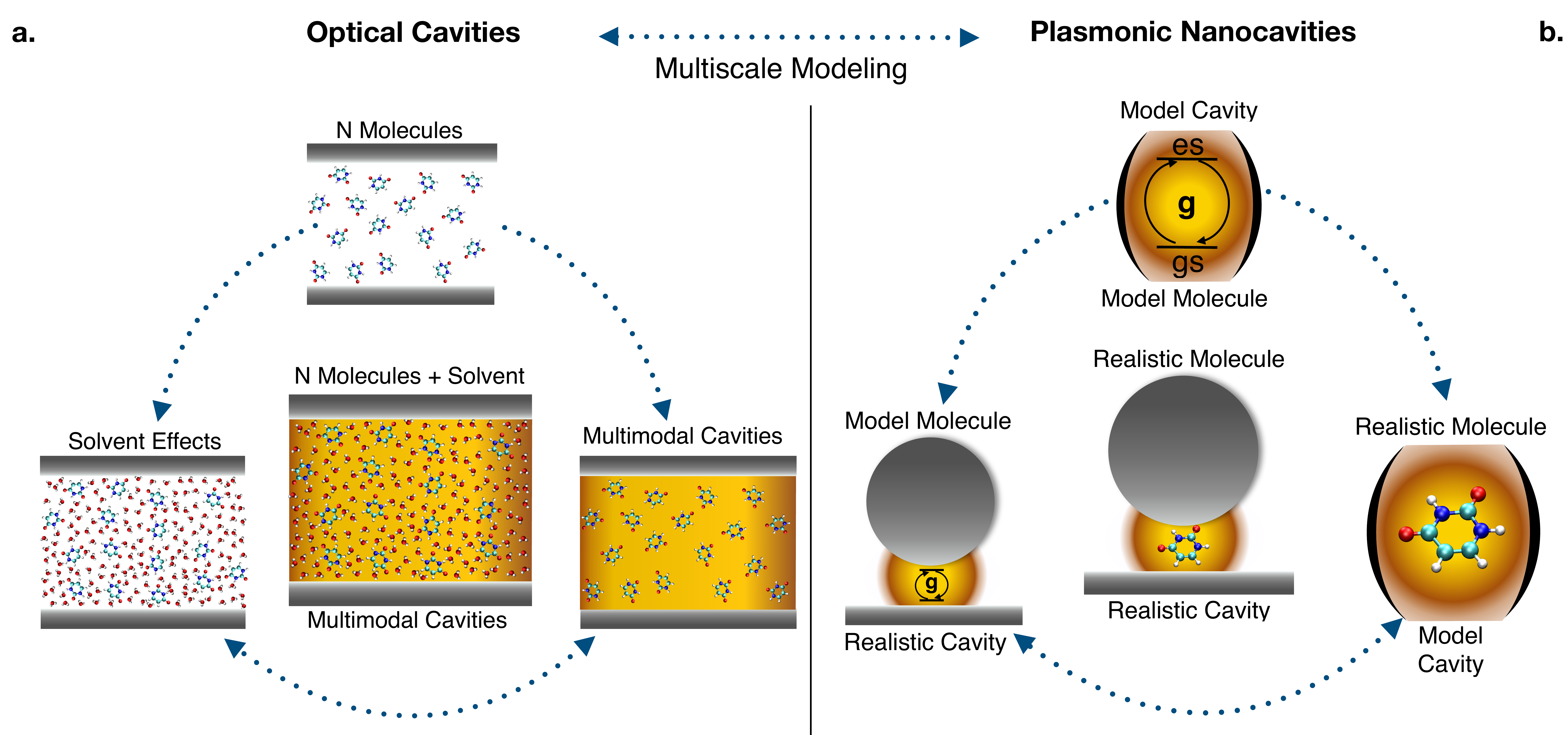}
	\caption{\textbf{Polaritonic chemistry as a multiscale problem}
	\textbf{(a)} The challenges in modeling polaritonic chemistry in photonic
	cavities involve the description of large ensemble of molecules collectively
	coupled, all embedded in a complex chemical environment.
	\textbf{(b)} The challenges in achieving a detailed description in plasmonic
	nanocavities involve the accurate modeling of the plasmonic inhomogeneous
	electromagnetic field, to be interfaced with an accurate quantum-chemical
	treatment of (relatively few) molecules.}
	\label{fig:1}
\end{figure}

In this section, we provide an overview of typical experimental setups that have
been explored on the road towards polaritonic devices\cite{Sanvitto2016} to
control chemistry. Organic (often dye) molecules are commonly used, which have
excitation energies of a few eV and linewidths about an order of magnitude
smaller (at room temperature). Most experiments can be categorized into one of
two distinct groups that are distinguished by the photonic platform and the
number of involved molecules (see \autoref{fig:1}). The first are optical
cavities, most often formed by planar mirrors (Fabry-Pérot cavities). The cavity
modes are then standing waves with characteristic dimensions similar to the
free-space wavelength. In such systems, strong coupling is achieved with
macroscopic numbers of molecules\cite{Lidzey1998, Lidzey1999, Schwartz2011}, as
depicted in \autoref{fig:1}(a). The relatively large size of such cavities means
that fabrication is not too challenging, and allows the use of liquid
samples\cite{Bahsoun2018}.

The second type of cavities are sub-wavelength plasmonic (i.e., metallic)
cavities where the ``light'' modes are characterized by collective oscillations
of the electrons in the structure, which permits the concentration of one
quantum of excitation to spatial scales far below the free-space wavelength.
While such systems are often referred to as (nano-)cavities for simplicity, a
more physically accurate nomenclature is ``resonator'' or ``antenna''. Effective
mode volumes (roughly proportional to the physical volume occupied by the EM
mode) can reach below $100$~nm$^3$\cite{Chikkaraddy2016} and possibly even down
to $\approx 1$~nm$^3$\cite{Urbieta2018, Carnegie2018, Wu2021, Li2021Bright}.
Such setups, depicted in \autoref{fig:1}(b), allow strong coupling to be reached
with a few molecules\cite{Zengin2015, Heintz2021} or even a single
emitter\cite{Chikkaraddy2016, Santhosh2016, Leng2018, Ciraci2019}.

Despite a difference of many orders of magnitude in effective volume of the
modes and the number of involved molecules, typical Rabi splittings
(corresponding to the energy difference between the two polariton modes formed
when a molecular transition and a cavity mode are on resonance) in both systems
are comparable and range from $\Omega_R \approx 100$~meV up to more than an
eV\cite{Gambino2014, Schwartz2011, Eizner2018}. At first sight, it might seem
somewhat surprising that such physically different systems lead to similar
effective coupling strengths, but this is actually straightforward to
understand. To do so, we treat a simplified model of $N$ identical two-level
molecules (where only the lowest two electronic states are taken into account
and rovibrational motion is ignored) that are all coupled identically to a
single EM mode, such that the space-dependent electric field profile is ignored.
In that situation, the Rabi splitting is given by\cite{Garraway2011}
\begin{equation}
    \Omega_R = 2\sqrt{N} \bm{\mu} \cdot \mb{E}\qquad\mathrm{with}\qquad |\mb{E}| = \sqrt{\frac{\hbar\omega}{2\varepsilon_0 \varepsilon_r \Veff}},
\end{equation}
where $\bm{\mu}$ is the molecular transition dipole moment, $\Veff$ is the
effective mode volume of the confined EM field, $\varepsilon_0$ is the vacuum
permittivity, and $\varepsilon_r$ is the relative background permittivity of the
molecular material. The result is that $\Omega_R \propto \sqrt{\mu^2 N/\Veff}$,
which implies that the Rabi splitting is proportional to the dipole density of
the molecular material, but does not depend separately on the absolute number of
molecules or volume of the cavity mode. In other words, large cavities give the
same Rabi splitting as small ones because the per-molecule coupling decreases
but they can be filled with more molecules. A more detailed study shows that the
Rabi splitting is proportional to the square of the dipole density times a
scalar filling factor (ranging between $0$ and $1$) that measures the fraction
of the photonic mode that is filled with the molecular
material\cite{Abujetas2019, *Abujetas2019Erratum}. The Rabi splitting can also
be rewritten in terms of the amplitude of the molecular transition obtained when
expressing the dielectric function of the molecular material using a Lorentz
oscillator model, and can thus be calculated from directly measurable
macroscopic quantities. The maximum splitting that can be reached for a given
material turns out to be the well-known value obtained for bulk
polaritons\cite{Hopfield1958}, and is independent of cavity
geometry\cite{Abujetas2019, *Abujetas2019Erratum, Canales2021,
Barra-Burillo2021}.

While the available Rabi splittings are similar, the two types of setups have
complementary strengths and weaknesses, and thus serve quite different uses. As
commented above, optical microcavities are characterized by large mode volumes
and thus require macroscopic numbers of molecules to achieve strong coupling,
with typical values ranging from $10^6$ to $10^{10}$ molecules per cavity
mode\cite{DelPino2015Quantum, Eizner2019, Arnardottir2020} at optical
frequencies, and even more at IR frequencies under vibrational strong coupling.
The polaritonic modes are then delocalized over many molecules, giving rise to
collective effects and effective long-range interactions between spatially
separated molecules\cite{Schwartz2011, Hutchison2012, Galego2016, Galego2017,
Munkhbat2018, Peters2019, Coles2014, Georgiou2018Control, Zhong2017,
Garcia-Vidal2017, Du2018Theory, Saez-Blazquez2018Organic, Groenhof2019,
Saez-Blazquez2019, Tichauer2021, Rozenman2018, Satapathy2021}. While there is a
wide range of designs that have been developed for optical light
confinement\cite{Vahala2003, Zhu2020, Scott2020, Hu2018Experimental},
experiments in polaritonic chemistry have almost exclusively used Fabry-Pérot
cavities consisting of two planar mirrors. The mirrors are typically either made
of metal or from distributed Bragg reflectors (DBRs, alternating layers of
dielectric materials with different refractive indices). Metal mirrors are
easier to fabricate, but at optical frequencies lead to quite lossy cavity modes
with low quality factors ($Q \approx 10$), where $Q=\omega_c/\kappa$ is the
ratio between the cavity mode frequency $\omega_c$ and its decay rate $\kappa$,
and corresponding lifetimes $\tau = 1/\kappa$ on the order of $10$~fs. In
contrast, DBR mirrors can be fabricated with relatively high reflectivity and
low losses, giving quality factors on the order of $Q=1000$ and cavity mode
lifetimes on the picosecond scale.

Subwavelength nanocavities also feature a very large flexibility in the design,
with the field confinement being tunable through the the size and shape of the
plasmonic platform\cite{Hugall2018}. The large confinement typically leads to a
strongly inhomogeneous EM field profile\cite{Neuman2018Coupling,
Cuartero-Gonzalez2018, Cuartero-Gonzalez2020Dipolar}, in particular, when atomic
extrusions form so-called picocavities\cite{Benz2016, Carnegie2018}. This makes
accurate placement of the emitters crucial, which can for instance be achieved
through the use of DNA origami\cite{Acuna2012, Chikkaraddy2016, Ojambati2019}.
Due to the intrinsic losses present in metals\cite{Khurgin2015}, plasmonic
nanocavity modes are limited to short lifetimes (typically below
$10$~fs)\cite{Matsuzaki2021}, such that most dynamics become dominated by
ultrafast radiative and nonradiative decay. While this poses a challenge for
polaritonic chemistry approaches that rely on dynamics in the excited state,
these fast losses can also be exploited to open up additional relaxation
channels that can be beneficial for the desired application, such as
photoprotection\cite{Galego2016, Felicetti2020, Davidsson2020Simulating,
Antoniou2020}, suppression of undesired side reactions\cite{Munkhbat2018},
opening of new reaction channels\cite{Torres-Sanchez2021}, sensing
applications\cite{Maccaferri2021}, and imaging techniques for ultrafast
processes\cite{Silva2020}.

\section{Theoretical approaches and challenges}

As the above discussion shows, theoretical approaches aimed at describing the
rich phenomenology of molecules strongly coupled to confined EM modes encounter
an inherently multiscale problem, with distinct challenges depending on which
type of situation is to be treated: large ensembles of molecules with collective
effects and long-range phenomena (in optical microcavities), or few molecules
interacting with a complex, highly lossy and inhomogeneous electromagnetic
environment (plasmonic nanocavities). In this section, we discuss the principal
aspects and approaches that have been developed over the past few years to treat
such systems.

The ``correct'' theory for describing molecules is nonrelativistic quantum
electrodynamics (QED)\cite{Cohen-Tannoudji1997, Cohen-Tannoudji1998}, which
describes the interaction between charged point particles (electrons and nuclei)
and EM fields. In general, the coupled Hamiltonian (in Coulomb gauge) can be
written as
\begin{equation}
    \hat H = \Hch + \HEM + \HcE,
\end{equation}
where $\Hch$ describes the the kinetic energies and Coulomb interactions of the
charged particles, $\HEM$ describes the radiative (transversal) EM field modes
(which are harmonic oscillators), and $\HcE$ describes the interactions between
charges and EM modes. In free space (and in the absence of external driving
fields), the interaction between light and matter is weak and its main effect is
the radiative decay of excited states due to the spontaneous emission of photons
(excitations of the free-space EM field). The standard approach of quantum
chemistry is thus to only treat $\Hch$ explicitly to obtain the approximate
molecular energy structure (exact solutions are only possible for the very
smallest molecules), and to either ignore spontaneous emission completely (when
only short-time dynamics are of interest), or to treat it perturbatively.
Typical spontaneous emission lifetimes for good molecular emitters (i.e.,
molecules with large transition dipole moments, $\mu \sim 10$~Debye) at optical
frequencies are on the order of a few nanoseconds, with some J-aggregates (where
a collective excitation is distributed over $N$ monomers) reaching down to tens
of picoseconds at cryogenic temperatures\cite{Fidder1990,Moll1995}. This is slow
compared to vibrational relaxation and thermalization, which typically happen on
sub-picosecond to few-picosecond scales\cite{May2011}. In cavities, the role of
the EM field becomes more relevant and the assumption that $\Hch$ can be treated
separately breaks down when the light-matter interaction becomes strong enough.
It then becomes necessary to also treat $\HEM$ and $\HcE$ explicitly to obtain
the correct energies and states of the coupled system. Therein lies the rub of
polaritonic chemistry.

Before turning to more practical considerations, we point out that in the above
statement about the importance of EM modes in cavities, we have silently changed
the concepts we are using by pretending that a ``cavity'' is an abstract way of
changing the EM mode Hamiltonian. In line with this useful lie, cavity modes are
often described as arising from applying boundary conditions to the EM field
modes. However, in reality, any cavity is a \emph{material} system, i.e., a
collection of charged particles (such as mirrors or plasmonic nanoantennas) that
are arranged so as to influence the EM field modes and to achieve the desired
properties. It is thus more correct to perform a repartitioning $\hat{H}$, with
the parts of $\Hch$ and $\HcE$ describing the cavity material and its
interaction with the EM field being grouped with $\HEM$ and forming a new
``cavity'' Hamiltonian $\Hcav$, such that 
\begin{equation}\label{eq:Hmolcav}
	\hat H = \Hmol + \Hcav + \Hmc,
\end{equation}
where $\Hmol$ is now only the molecule (or any other material system) that will
be treated in detail, while $\Hcav$ describes the combined excitations of the
coupled cavity material and free-space EM modes. Under the assumption that the
cavity material can be treated through linear response, diagonalizing $\Hcav$ is
equivalent to solving the macroscopic Maxwell equations(see
Ref.~\nocite{Feist2020}\citenum{Feist2020} for an overview). It is in this sense
that $\Hcav$ is often said to describe the EM field, and its excitations are
called ``photons''. In particular, its eigenmodes keep being harmonic
oscillators. However, ignoring the simple fact that $\Hcav$ also includes
material response can have serious consequences and lead to misleading
conclusions. For example, plasmonic nanocavity modes mostly
correspond to material excitations (collective oscillations of the electrons in
the metal), and their interaction with the molecules are mostly mediated by
(longitudinal) Coulomb interactions, not by (transversal) free-space EM modes.
The Coulomb interaction is not affected by the Power-Zienau-Woolley
transformation and, in particular, gives an $\vec{E}\cdot\vec{d}$ interaction
even in minimal coupling, without any dipole-self-energy term\cite{Galego2019,
Feist2020}. The dipole-self-energy term should thus not be included when
treating a physical situation corresponding to a strongly subwavelength (e.g.,
plasmonic) nanocavity, which is the only available way to approach
single-molecule strong coupling. Results in the literature with single-molecule
strong coupling where the dipole self-energy term is included should therefore
be approached with care.

As mentioned above, when assuming linear response for the cavity material,
$\Hcav$ can be diagonalized as a collection of harmonic oscillators, just like
the free-space EM field. Formally, there is always a continuum of solutions
existing at any (positive) energy. In practice, this can often be reduced to an
effective description where only a single or a few ``cavity modes'' have to be
treated explicitly, although the coupling to the residual continuum means that
these cavity modes are generally resonances with finite (and possibly very
short) lifetimes\cite{Koenderink2010, Franke2019, Medina2021}.

After these general considerations, which are normally skipped over in the
literature (which has to be done with care, as discussed above), we have thus
finally arrived at the Hamiltonian that is often the starting point in the
literature on polaritonic chemistry. We now discuss available approaches for
solving the Hamiltonian \autoref{eq:Hmolcav} which describes three types of
degrees of freedom: electronic ($\mb{r}$), nuclear ($\mb{R}$), and photonic
($\mb{q}$). Depending on the level of description with which each of its terms
is treated, we can roughly categorize the numerous methods available in
literature by their level of realism, as sketched in \autoref{fig:1}. We note
that for consistency, we write the cavity modes using the ``position space''
degrees of freedom $\mb{q}$. The Hamiltonian of a cavity mode with frequency
$\omega_c$ is $H_\mathrm{mode} = \frac12 p_q^2 + \frac{\omega_c^2}{2} q^2$,
which can equally be expressed in terms of the ladder operators, $a =
\sqrt{\frac{\omega_c}{2}} \left(q + \frac{i}{\omega_c} p_q\right)$, giving
$H_\mathrm{mode} = \omega_c \left(a^\dagger a + \frac12\right)$. This form is
typically used in quantum optics as it allows a natural interpretation of the
operators $a$ and $a^\dagger$ as annihilating and creating a photon,
respectively.

\begin{figure}[ht!]
	\includegraphics[width=\linewidth]{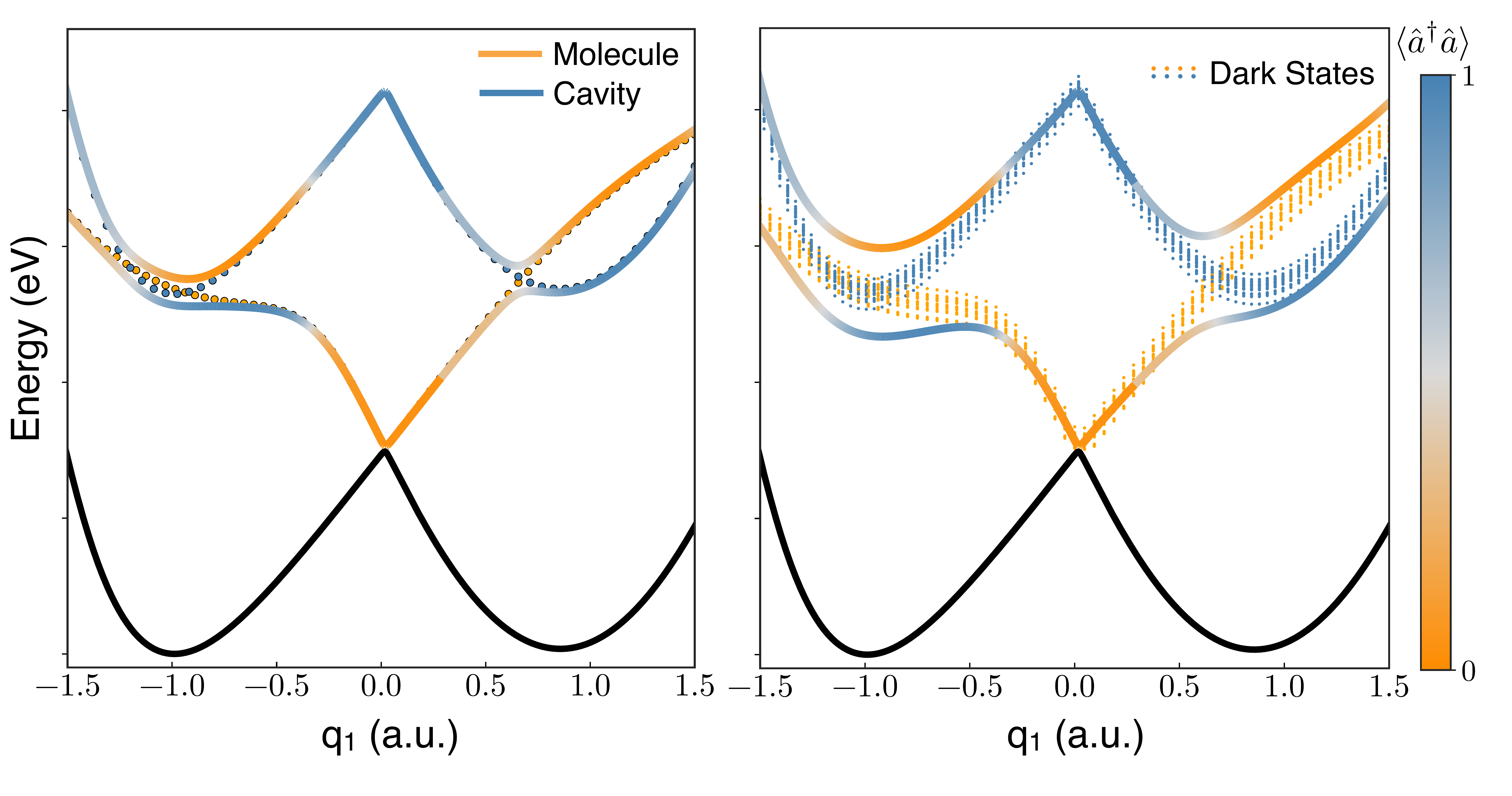}
	\caption{\textbf{Polaritonic Potential Energy Surfaces} \textbf{(a)} Case of
	a single molecule strongly coupled with light, where coupling between the
	cavity (blue dotted) and the molecule (orange dotted) states couple to
	originate polaritons. \textbf{(b)} Case of a molecular ensemble (N=50), where
	a manifold of dark states emerges.}
	\label{fig:2}
\end{figure}

When treating a system described by the Hamiltonian~(\ref{eq:Hmolcav}), it can
be helpful to factorize the time-dependent wavefunction
$\Psi(\mb{r},\mb{R},\mb{q},t)$ using a Born-Huang expansion, where slow and fast
degrees of freedom are separated. In electronic strong coupling, which we focus
on here, the cavity mode frequencies are (close to) resonant with electronic
transitions, and the dynamics of electrons and cavity modes are thus comparably
fast, making it natural to group them together\cite{Galego2015, Luk2017,
Fregoni2020Photochemistry}
\begin{equation}\label{eq:bh2}
	\Psi(\mb{r},\mb{R},\mb{q},t) = \sum_k \chi_k(\mb{R},t)\phi_k(\mb{r},\mb{q};\mb{R}),
\end{equation}
Here, the states $\phi_k(\mb{r},\mb{q};\mb{R})$ are the eigenstates of the
Hamiltonian without the nuclear kinetic energy. They are mixed photonic and
electronic (polaritonic) states that parametrically depend on the nuclear
coordinates, with the associated energies being polaritonic potential energy
surfaces (PoPES)\cite{Feist2018}. Potential energy surfaces as a tool are
extensively used to simulate and predict the properties and outcomes of
photochemical reactions. As such, the adaptation of this tool to polaritonic
chemistry can describe how the energy landscape---and consequently the
reactivity---is modified when molecules are brought into strong coupling. The
PoPES also give information about whether the excitation on a given
surface is more photon- or more exciton-like, i.e., whether the energy is stored
in the cavity or in the molecule, as sketched in \autoref{fig:2}a. Typically, a
single PoPES will gradually change its character as a function of nuclear
coordinate. This can lead to periodic transfer of energy between the molecule
and cavity due to nuclear motion in a process that is completely distinct from
conventional vacuum Rabi oscillations and could allow, e.g., following the
nuclear wave packet motion in time\cite{Silva2020}.

We note that it is also possible to use exact factorization methods for
analyzing the cavity-induced molecular dynamics\cite{Lacombe2019}, or to group
the photonic and nuclear coordinates together such that electronic states
parametrically depend on the photonic and nuclear coordinates $\mb{q},\mb{R}$,
leading to the so-called cavity Born-Oppenheimer (CBO)
approximation\cite{Flick2017Cavity}. This approach is especially powerful in the
regime of vibrational strong coupling (VSC), where nuclear motion and photonic
dynamics are comparably fast and the dynamics usually takes places on the lowest
(ground) electronic state\cite{Galego2019,Fischer2021}. In contrast, it is not
ideal for describing photochemical processes in strong coupling, as the $n+p$
representation requires the propagation of the full quantum nuclear+photon
wavefunction $\chi_k(\mb{R},\mb{q},t)$.

In order to obtain the PoPES and the nonadiabatic couplings between them, it is
thus necessary to solve the coupled electron-photon Hamiltonian. There are two
main strategies that have been followed to achieve this, both of which are
formally exact and fully \emph{ab initio}, but have different strengths and
weaknesses. The first is conceptually comparable to a configuration interaction
(CI) approach where the Hamiltonian is first diagonalized without including the
light-matter interaction, and the eigenbasis of the uncoupled Hamiltonian is
then used to express and diagonalize the full Hamiltonian. This approach has
several clear advantages. On the one hand, it is quite straightforward to
implement, as it allows to use any of the methods in the toolbox of standard
quantum chemistry (QC) to solve the molecular problem. If the light-matter
coupling is treated in the commonly used dipole (or long-wavelength)
approximation, only the electronic energies and (permanent and transition)
dipole moments have to be calculated. We note that permanent dipole moments are
often disregarded in the literature, which implicitly corresponds to assuming
that the permanent dipole moment is approximately independent of electronic
state and nuclear position, which is not necessarily a good approximation.
Higher-order light-matter couplings such as quadrupolar
interactions\cite{Cuartero-Gonzalez2018} can also be included if the quadrupole
moments are calculated. Second, it allows for an easy interpretation of the
resulting polaritonic states, as they are expressed as superpositions of the
physical eigenstates of the uncoupled system with well-defined properties.
Finally, the convergence of the approach can be tested by including successively
more electronic states, and is usually quite rapid, in particular when the
per-molecule coupling strength is not too large. In particular, it is often
sufficient to only include two electronic states (the ground and first excited
state). In the literature, a wide range of quantum chemistry (QC) methods have
been employed to provide the input for this CI-like treatment of polaritonic
chemistry, such as TDDFT\cite{Luk2017}, semiempirical
methods\cite{Fregoni2020Photochemistry}, MRCI\cite{Davidsson2020Atom} and
CASSCF\cite{Mony2021}.

The second strategy to treat light-matter coupling within the electron-photon
Hamiltonian relies on extending QC methods to directly include cavity modes in
their solution. The advantages of these approaches is that they are expected to
more easily capture changes in state wave functions that would require large
expansions in the polaritonic CI approach discussed above. This becomes
especially relevant when coupling strengths are large. Two notable developments
in this direction are QE-DFT\cite{Tokatly2013, Ruggenthaler2014} and
QED-CC\cite{Folkestad2020, Haugland2020}. The former is computationally cheap,
but inherits the intrinsic problems of density functional theory approaches
since all known exchange and correlation functionals correspond to severe
approximations\cite{Pellegrini2015, Flick2018Abinitio}. The latter offers a
robust but computationally expensive alternative. As mentioned above, the
strength of these approaches lies in the description of electronic-photonic
states that are not just superpositions of closely lying uncoupled states, which
happens for large enough coupling strengths. In the CI approach, convergence
then requires the calculation of an enormous number of excited states. It is
then at some point computationally cheaper and more straightforward to calculate
the ``new'' electronic-photonic states directly instead of using the uncoupled
states as the expansion basis. However, it should here be noted that
single-molecule changes usually depend on the single-molecule coupling strength
and are not collectively enhanced in many-molecule setups\cite{Galego2015,
Cwik2016, Pilar2020}. This effect is thus not expected to be present in such
systems, and few-state expansions should work well. In contrast, for the largest
single-molecule coupling strengths available (in plasmonic nanocavities with
gaps on the order of $1$~nm\cite{Chikkaraddy2016}), treating the cavity mode as
a lossless photonic mode and neglecting the atomistic structure of the plasmonic
nanocavity are both severe approximations\cite{Zhang2014,Rossi2019}.

Once the method to obtain the polaritonic (electronic-photonic) structure of a
given problem has been chosen, some way to treat the nuclear motion has to be
included. The cheapest method is to not do any nuclear dynamics, i.e., to simply
analyze the obtained PoPES\@. This can already provide significant insight about
the possible changes in the system response due to strong coupling, but of
course precludes any quantitative insight. Going beyond this, semiclassical
methods based on surface hopping are powerful tools\cite{Luk2017, Fregoni2018,
Fregoni2020Strong, Antoniou2020}, as they can qualitatively describe a large
number of nuclear degrees of freedom when a relatively small number of excited
state is involved. As such, they are best exploited to describe one to a small
number of molecules, as the algorithm fails at grasping collective effects even
in the more refined implementations\cite{Granucci2010, Jaeger2012, Plasser2019,
Kossoski2020}. The failure is due to the inaccurate evaluation of transition
probabilities in the presence of many quasi-degenerate
states\cite{Granucci2007}, which is exactly the case typically encountered when
many molecules couple to a single cavity mode\cite{Feist2018,
Vendrell2018Collective}. An additional problem for the current implementations
of semiclassical algorithms that may be potentially hindering to polaritonic
chemistry is the incapacity of describing tunneling through potential energy
surfaces. A palliative solution to this problem comes from partially including
the nuclear quantum effects in the semiclassical simulations, for example with
the ring polymer technique\cite{Shushkov2012, Shakib2017}. One big advantage
that semiclassical techniques offers are that it becomes easier to include more
of the environmental complexity, such as atomistic descriptions of the
solvent\cite{Fregoni2020Strong} and chemical environment\cite{Luk2017}, achieved
by including the electrostatic interactions between classical MM charges and the
QM charge density (electrostatic embedding). Furthermore, trajectory-based
approaches\cite{Persico2014} allow the straightforward inclusion of cavity
losses via quantum jump algorithms\cite{Dalibard1992, Molmer1993} in the
framework of stochastic Schrödinger equation (SSE)\cite{Biele2012, Coccia2018,
Coccia2020} and non-hermitian formulations\cite{Gao2017, Antoniou2020}.

As a counterpart to semiclassical techniques for the treatment of nuclear
motion, quantum wavepacket dynamics can provide highly accurate results for a
restricted number of degrees of freedom, with the drawback of much larger
computational cost. For low-dimensional model problems, direct grid-based
methods are relatively straightforward to implement and provide accurate
solutions\cite{Galego2015, Kowalewski2016Cavity, Bennett2016, Galego2016}. For
high-dimensional nuclear wave functions, the method of choice is the
Multiconfigurational Time-Dependent Hartree (MCTDH)
algorithm\cite{Vendrell2018Coherent, Ulusoy2020}, possibly in its multilayer
implementation\cite{Wang2015}. When potential surfaces can be approximated as
harmonic oscillators, tensor network approaches are another powerful way to
perform full quantum dynamics\cite{DelPino2018Dynamics,Zhao2020} As a hallmark
feature, methods relying on wavepacket propagations guarantee a correct dynamics
of the nuclear wavepacket at both electronic and polaritonic avoided crossings,
conical intersections, and seams between the PoPESs, including a correct decay
of nuclear coherence without needing to resort to artificial corrections as in
the semiclassical methods. Secondly, its propagation allows to exactly include
decay channels in the dynamics, either through effective non-Hermitian
Hamiltonians\cite{Silva2020, Antoniou2020, Felicetti2020} that are exact when
the dynamics after decay are not of interest, or by direct solution of a
Lindblad-style master equation\cite{Torres-Sanchez2021,
Davidsson2020Simulating}. This feature is particularly advantageous when the
polaritonic relaxation involves multiple polaritonic states and the decay
mechanism is an interplay between radiative and non-radiative transitions. These
characteristics make wavepacket dynamics an excellent investigation tool to
explore the effect of cavity losses or the role of strong coupling on conical
intersections\cite{Gu2020Manipulating, Szidarovszky2018, Fabri2021}.

The propagation schemes for nuclei have proven instrumental in surveying new
effects and predicting new applications when few molecules are involved. Among
them, we count the suppression/enhancement of photoisomerization reactions,
photoprotection/photostability of organic chromophores\cite{Galego2016,
Felicetti2020, Davidsson2020Simulating, Gudem2021, Antoniou2020},
photodissociation\cite{Kowalewski2016Cavity, Bennett2016,
Kowalewski2016Nonadiabatic, Torres-Sanchez2021}, reverse intersystem crossing
(RISC)\cite{Eizner2019,Yu2021Barrier-free}.

A common approximation in the methods discussed above is to rely on a single
cavity mode. An extension to the case of multimodal cavities has been
implemented only recently\cite{Tichauer2021}. Furthermore, only few approaches
have tried to combine a quantum chemical description of the molecule with a
realistic nanophotonic setup. These approaches rely on the quantization of the
electromagnetic environment via different approaches\cite{Felicetti2020,
Fregoni2021}. It is an open question and important challenge to understand
whether such approaches are valid in the limit of atomistic resolution that is
approached in recent experiments in nanoplasmonic gap
cavities\cite{Zhang2013Chemical, Doppagne2020} even though they rely on continuum
descriptions of the cavity (plasmonic) medium. There are encouraging indications
that this is possible\cite{Urbieta2018}. As such, these methods will be
potentially able to guide the investigation of polaritonic chemistry in setups
confining the electromagnetic field at sub-nanometric volumes, such as
picocavities\cite{Benz2016, Carnegie2018}.

Despite the accurate level of description reached for strong coupling in
few-molecule problems, the modeling of polaritonic reactions meets an intrinsic
problem when trying to describe large ensembles. Most of the polaritonic
chemistry experiments are performed in microcavities, where up to $N=10^{10}$
emitters are involved. In principle, the PoPESs in such a setup are $N\times
N_m$-dimensional, where $N_m$ is the number of nuclear degrees of freedom
required to describe a single emitter molecule (possibly including the chemical
environment). If the molecules were decoupled, the strategy would be to treat a
restricted number of molecules (one-to-few) via quantum chemistry methods,
including the chemical environment molecules (solvent or protein scaffolds)
atomistically (QM/MM techniques) or as a continuum medium (PCM
techniques)\cite{Mennucci2019}. Instead, the strongly delocalized
electromagnetic field in the cavity opens up long-range interaction channels in
a disordered ensemble of molecules\cite{Sommer2021}. This makes it highly
challenging to infer photochemical properties of an ensemble of $N$ molecules
from the detailed study of a very restricted subset of it. To take into account
the large number of emitters, one approach is to use strongly simplified
molecular models, such as the Holstein model where each molecule is described by
two displaced harmonic oscillators describing nuclear motion in the electronic
ground and excited states. This allows including a large (few thousands) number
of molecules, coupled to the cavity with a Tavis-Cummings-like
model\cite{Herrera2020}. Despite its success in predicting long-range energy
transfer\cite{Saez-Blazquez2018Organic, Du2018Theory} and remote
catalysis\cite{Du2018Theory}, the exciton-based approaches present several
drawbacks. The most evident is that the non-atomistic description does not allow
to grasp structural rearrangements of molecules upon, e.g., charge transfers and
the associated chemical environment rearrangement. This can be included by
approaching the problem of collective effects using multiscale
techniques\cite{Luk2017, Groenhof2019, Li2021Collective, LiNitzan2021Cavity}.
The approach initially developed by Luk \emph{et al.}\cite{Luk2017} already
implements a QM/MM description of molecules in cavities, and has been extended
to a multimode cavity characterized by a 1D dispersion\cite{Tichauer2021}. Its
current implementation already supports a large number of both wavefunction and
density functional methods, interfaced with both surface hopping and Ehrenfest
dynamics\cite{Luk2017}. In the presence of many molecules and thus a large
manifold of closely spaced PoPES, Ehrenfest dynamics provide more robust results
compared to surface hopping approaches\cite{Groenhof2019}.

\begin{figure}[ht!]
	\includegraphics[width=\linewidth]{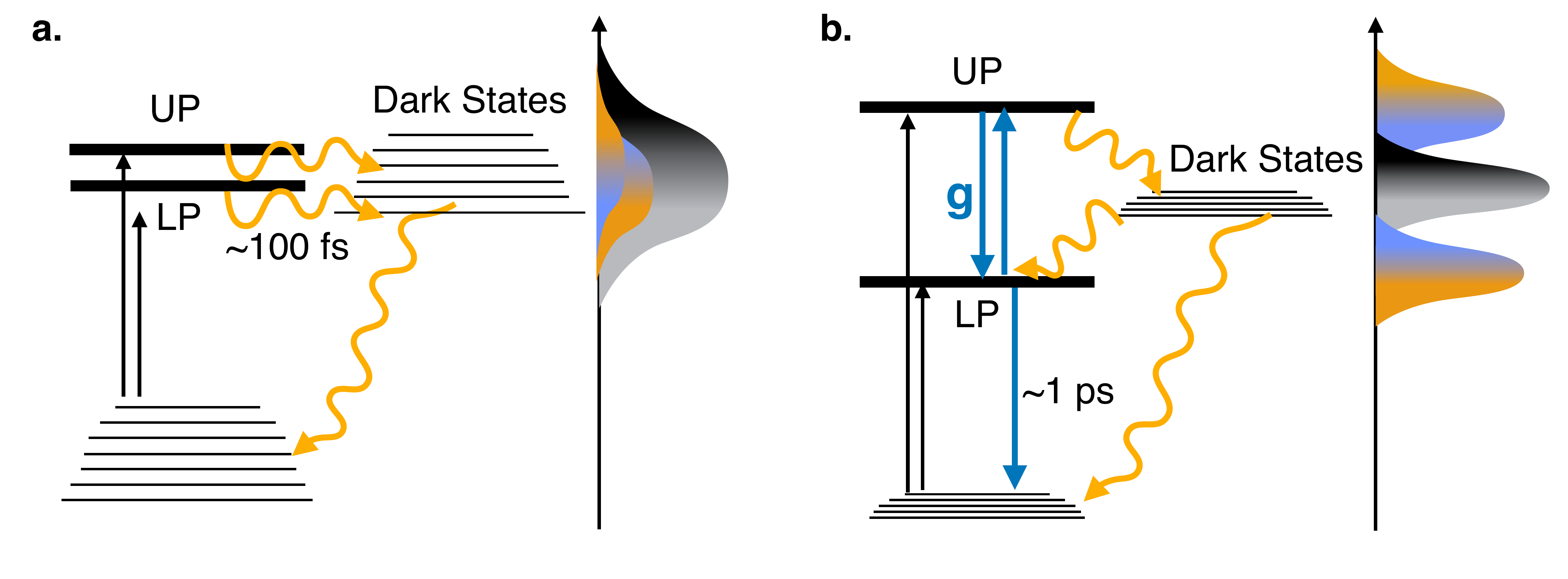}
	\caption{\textbf{Dominant processes in polaritonic systems.} \textbf{(a)}
	Ultrafast decoherence of the cavity excitation. The overlap between
	polaritonic bright and dark states funnels the wavepacket towards the dark
	manifold, where the wavepacket undergoes decoherence via non-radiative
	processes of the individual molecules. \textbf{(b)} Dominant processes occur
	from the lower polaritonic state, as the dark manifold are decoupled from
	the polaritonic states. This scheme implies a long-lived delocalized
	excitation, which can potentially result in a cavity-modified chemistry.}
	\label{fig:3}
\end{figure}

A prominent signature of the necessity to describe large ensembles is the
emergence of a dark state manifold when ensembles of molecules are coupled to a
cavity (\autoref{fig:2}(b)). Within the first excited subspace, there are $N+1$
states, each of them corresponding to a single excitation (either in one of the
$N$ molecules or in the cavity mode) of the global system from its ground state.
The so-called bright state is obtained when the molecular excitation is
delocalized over all the resonant molecules. This states couples ideally to the
cavity mode (with effective coupling enhanced by $\sqrt{N}$ over the
single-molecule one). The molecular bright state and the cavity mode couple to
form the typical upper and lower polariton modes. In a simple conceptual
picture, all the other orthogonal superpositions obtained by distributing a
single excitation over the molecules constitute the $N-1$ dark states manifold.
We note that this simple picture is only true in the case of perfectly
degenerate two-level emitters\cite{Houdre1996}, but it provides a convenient
framework to think about the states in the system. In particular, when the
molecules are not identical (or the nuclear configurations are distinct, even
for nominally identical molecules), the dark states are not fully dark and
provide residual light absorption and emission. While it is conceptually common
to think about the dark states as states in which the excitation is localized on
individual molecules, it has been shown that the dark state manifold inherits
some of the delocalized polaritonic properties\cite{Gonzalez-Ballestero2016Uncoupled,
DelPo2020}. Still, the energy distribution of dark states closely matches the
absorption spectrum of the bare molecules\cite{Coles2011, Groenhof2019}.
Furthermore, the potential energy landscape of each of these states
(\autoref{fig:2}(b)) looks quite similar to the collective ground state of the
isolated molecular ensemble. The role of dark states in polaritonic processes is
then strongly dependent on the specifics of the system: when the dark states
manifold embeds (strongly overlaps with) polaritonic states, which in particular
happens for broadband absorbers\cite{Mony2021, DelPo2021}, the polaritons
dephase into cavity-free superpositions of states in the dark manifold. This
ultrafast loss of coherence to the dark states can become the dominant decay
process for polaritons\cite{Groenhof2019}, see (\autoref{fig:3}a), resulting in
reactivity essentially equal to that of isolated molecules\cite{Mony2021}. Put
in another way, if we want to ensure that photochemical reactions can
efficiently take place on the polaritonic potential energy surfaces, the Rabi
splitting should be larger than the molecular absorption band. We note that the
lifetime of the polaritons is not limited by the molecular absorption bandwidth
since the latter is dominated by the spread of molecular excitation energies,
not by the intrinsic lifetime of molecular excitations\cite{Houdre1996}. This
implies that there is no reason to ``match'' the cavity bandwidth to the
molecular absorption band, and indeed, when the polaritons do not overlap with
the dark states, the dominant decay process becomes radiative decay from the
lower polariton (\autoref{fig:3}b). This occurs at roughly half the bare-cavity
decay rate (which can translate to lifetimes from the few-femtosecond to
picosecond range), and can give linewidths much smaller than the bare molecular
one\cite{Gambino2014}. Such decay times are comparable to those of several
photochemical reactions\cite{Olivucci2005}, confirming the possibility to
influence photochemistry with polaritons.

\section{Conclusions \& Outlook}

Over the past years, polaritonic chemistry has developed into a vibrant field
that is drawing increasing attention both from the experimental and theoretical
communities. It holds the promise of providing an approach to control
(photo)chemical reactions that is completely distinct from traditional ones, and
in particular does not rely on the external input of energy apart from
absorption of single photons. The theoretical description of these processes
faces many challenges due to the inherently multiscale nature of the problem,
with unique challenges arising in each of the two distinct types of common
experimental setups. In wavelength-scale optical cavities, the macroscopic
number of participating molecules a priori prevents a full representation of
experimental reality in the theoretical approaches, as the sheer number of
degrees of freedom of the problem poses serious challenges even to semiclassical
approaches. Furthermore, there is usually a continuum of EM modes that has to be
taken into account for obtaining a complete picture. While experimentally much
simpler to construct than nanoplasmonic resonators requiring (sub)nanometric
precision, the theoretical treatment of cavity-modified molecular reactions in
wavelength-scale optical cavities thus faces a plethora of challenges and will
require the judicious use of appropriate approximations. 

In subwavelength cavities with single- or few-molecule strong coupling, accurate
descriptions are challenged by the large loss rates, the complex nature of the
EM field modes, and the importance of atomistic details in the material
structures providing the cavity modes. One way forward here will be given by
methods able to quantize the plasmonic electromagnetic field in arbitrary
material structures\cite{Franke2019, Medina2021, Fregoni2021}, and their
interface with quantum chemistry methods and non-adiabatic dynamics techniques
to account for the molecular reactivity. Going further, the inclusion of quantum
effects such as tunneling at the nanoparticle-molecule interface calls for a
multiscale layered technique, where the interface has to be described at a
quantum-mechanical atomistic level while still taking into account the global EM
modes and plasmonic excitations.

In addition to methodological challenges, there are also significant
experimental and conceptual obstacles to overcome on the path towards actual
devices based on the concepts of polaritonic chemistry. As an example,
strategies to either exploit or minimize losses are required, particularly in
subwavelength plasmonic cavities. There, the capability to reach longer
lifetimes would open up new intriguing phenomena taking place at the picosecond
timescale. One promising approach here could be provided by hybrid
metallodielectric cavities(see Ref.~\citenum{Medina2021} and references
therein), in which plasmonic excitations are hybridized with long-lived optical
cavity modes, allowing to control the tradeoff between strong field confinement
and material losses in metals. Another possibility that has not yet been
explored in this context are purely dielectric nanophotonic cavities designed to
achieve subwavelength field confinement while still largely avoiding
losses\cite{Hu2018Experimental}. In parallel, it remains to be seen whether the
use of atomic-scale extrusions (leading to picocavities) can enable control over
chemical reactions on the single-molecule level, possibly even with subnanometer
precision.

As in many previous cases of theoretical investigation, this search for
theoretical and numerical frameworks able to accurately describe the physical
and chemical process emerging in polaritonic chemistry at very different scales
will not only lead to a better understanding of the fundamental mechanisms
involved in the current experimental setups and guide the exploration of new
reliable platforms, but will also open new avenues for research in polaritonic
chemistry and related areas that we cannot foresee at this stage.

\section*{Acknowledgements}
This work has been funded by the European Research Council through Grant
ERC-2016-StG-714870 and by the Spanish Ministry for Science, Innovation, and
Universities -- Agencia Estatal de Investigación through grants
RTI2018-099737-B-I00, PCI2018-093145 (through the QuantERA program of the
European Commission), and CEX2018-000805-M (through the María de Maeztu program
for Units of Excellence in R\&D). We also acknowledge financial support from the
Proyecto Sinérgico CAM 2020 Y2020/TCS-6545 (NanoQuCo-CM).

\end{spacing}

\singlespacing{}
\bibliography{biblio}

\end{document}